\documentclass[namedreferences]{SolarPhysics}

\usepackage{graphicx}

\usepackage[optionalrh]{spr-sola-addons}
\usepackage{color}           
\usepackage{url}             

\begin{document}
\begin{article}

\begin{opening}

\title{Outflows at the Edges of an Active Region in a Coronal Hole: A Signature of Active Region Expansion?}

\author{M.~J.~\surname{Murray}$^{1}$\sep
        D.~\surname{Baker}$^{1}$\sep
        L.~\surname{van Driel-Gesztelyi}$^{1,2,3}$\sep
        J.~\surname{Sun}$^{1}$
       }
\runningauthor{M.~J. Murray et al.}
\runningtitle{Outflows at the edges of an active region in a coronal hole}

\institute{$^{1}$ University College London, Mullard Space Science Laboratory, Holmbury St. Mary, Dorking, Surrey, RH5 6NT, U.K.\\
              $^{2}$ Observatoire de Paris, LESIA, FRE 2461 (CNRS), F-92195 Meudon Principal Cedex, France\\
              $^{3}$ Konkoly Observatory of the Hungarian Academy of Sciences, Budapest, Hungary
             }

\begin{abstract}
Outflows of plasma at the edges of active regions surrounded by quiet Sun are now a common observation with the \emph{Hinode} satellite. While there is observational evidence to suggest that the outflows are originating in the magnetic field surrounding the active regions, there is no conclusive evidence that reveals how they are driven. Motivated by observations of outflows at the periphery of a mature active region embedded in a coronal hole, we have used a three-dimensional simulation to emulate the active region's development in order to investigate the origin and driver of these outflows. We find outflows are accelerated from a site in the coronal hole magnetic field immediately surrounding the active region and are channelled along the coronal hole field as they rise through the atmosphere. The plasma is accelerated simply as a result of the active region expanding horizontally as it develops. Many of the characteristics of the outflows generated in the simulation are consistent with those of observed outflows: velocities up to $45\ \mathrm{km}\ \mathrm{s}^{-1}$, properties akin to the coronal hole, proximity to the active region's draining loops, expansion with height, and projection over monopolar photospheric magnetic concentrations. Although the horizontal expansion occurs as a consequence of the active region's development in the simulation, expansion is also a general feature of established active regions. Hence, it is entirely possible and plausible that the expansion acceleration mechanism displayed in the simulation is occurring in active regions on the Sun and, in addition to reconnection, is driving the outflows observed at their edges. 
\end{abstract}
\end{opening}


\section{Introduction}\label{Sec:Intro}
Outflows at the edges of active regions (ARs) surrounded by quiet Sun have occasionally been observed with TRACE \cite{Winebarger01}. However, one of the major discoveries made by the \emph{Hinode} satellite \cite{Kosugi07} is that peripheral outflows appear to be a universal feature of ARs surrounded by quiet Sun. The X-Ray Telescope (XRT) \cite{Golub07} has provided visual evidence of the outflows \cite{Sakao07}, while the Extreme-ultraviolet Imaging Spectrometer (EIS) \cite{Culhane07} has provided velocity evidence of the outflows, derived from Doppler shifts. In contrast to previous spectrometers, such as SOHO/Coronal Diagnostic Spectrometer (CDS), EIS is able to see the outflows at the edges of ARs due to its measurement of spectral lines at higher temperatures and greater sensitivity and spectral resolution.

The blue-shifted Doppler flows, or outflows, are concentrated at the boundary between the hot closed loops of the AR and surrounding cooler long loops \cite{DelZanna08}. The AR loops are themselves red-shifted as plasma drains down them into the lower atmosphere and, thus, a sharp gradient in red and blue shifts exists at the boundaries between the closed loops and longer loops \cite{Harra08,Marsch08}. The line-of-sight velocities of the outflows have magnitudes of $10$ to $50\ \mathrm{km}\ \mathrm{s}^{-1}$, with magnetically stronger ARs being associated with faster outflows \cite{Doschek08}. The outflows have been observed to persist for a number of days with little change \cite{Sakao07,DelZanna08,Doschek08,Marsch08} and, as yet, observations showing the onset or halting of the outflows have not been reported.

Outflows have a defined spatial structure that is related to temperature. Analysis of AR 10942 by Baker \textit{et al.} \shortcite{Baker09} identifies that the same outflow is detected in a number of EIS's spectral lines, indicating that the outflowing plasma has a range of temperatures. More specifically, at temperatures of $630,000\ \mathrm{K}$ the outflow regions appear narrow and elongated while at hotter temperatures of $1-2\ \mathrm{MK}$ the outflow region has a larger spatial extent. Assuming a positive relationship between temperature and atmospheric height, this information points towards the outflows being accelerated from a small spatial region and fanning out as they rise, possibly following the natural divergence of the coronal magnetic field.

The properties of the outflow regions, such as temperature and density, are more representative of quiet Sun conditions than those of ARs \cite{Doschek08}. Electron densities in the outflow region of $7\times10^{8}\ \mathrm{cm}^{-3}$ derived from a density sensitive intensity ratio of Fe {\sc xii} lines by Doschek \textit{et al.} \shortcite{Doschek08} and $\approx3.2\times10^{9}\ \mathrm{cm}^{-3}$ estimated by Sakao \textit{et al.} \shortcite{Sakao07} are rather low for an active region. Similarly, the outflow regions have little radiance in contrast to active regions \cite{DelZanna08,Harra08}. Hence, it would appear that the outflows emanate from the region surrounding the AR as opposed to the AR itself. This theory is further supported by the fact that, in the absence of dynamic events, ARs change little as a result of the continuous outflows \cite{Marsch08}. However, a conundrum exists since it has been reported that the outflows occur over monopolar regions of significant field strength where AR loop footpoints are expected to be rooted rather than over magnetic conditions appropriate for the quiet Sun \cite{DelZanna08,Doschek08}.

Determining the driver of the outflows is exceedingly important since in some instances the outflowing plasma will be driven along magnetic field that extends into interplanetary space and, hence, the outflow regions have the possibility of being a source of the slow solar wind \cite{Sakao07,Doschek08,Harra08}. Several mechanisms have been tentatively proposed for generating the outflows including atmospheric evaporation as a result of heating during reconnection \cite{DelZanna08}, global plasma circulation \cite{Marsch08}, an impulsive coronal footpoint heating mechanism \cite{Hara08} and the reconnection of outer AR loops enabling expansion of the inner loops of the AR \cite{Harra08}. Reconstructions of the magnetic field topology reveal that the outflows originate from quasi-separatrix layers, locations where the field displays strong gradients of magnetic connectivity, thus lending further support for a reconnection-related mechanism driving the outflows \cite{Baker09}.

The aim of this paper is to investigate the mechanism(s) responsible for driving the outflowing plasma at the edges of ARs using a numerical simulation. The simulation is motivated by observations of outflows occurring at the periphery of an AR embedded in a coronal hole (CH). By simplifying the magnetic topology of the AR-CH complex, we have been able to identify a new simple driving mechanism for the observed outflows that requires no reconnection, namely horizontal AR expansion. The motivating AR-CH observations are introduced in Section~\ref{Sec:Obs}, the numerical method is described in Section ~\ref{Sec:Sim}, numerical results are presented in Section ~\ref{Sec:Res}. These are followed by discussions in Section ~\ref{Sec:Dis} and conclusions in Section ~\ref{Sec:Con}.

\section{Motivating Observations}\label{Sec:Obs}
\begin{figure}
\includegraphics[width=1.0\textwidth]{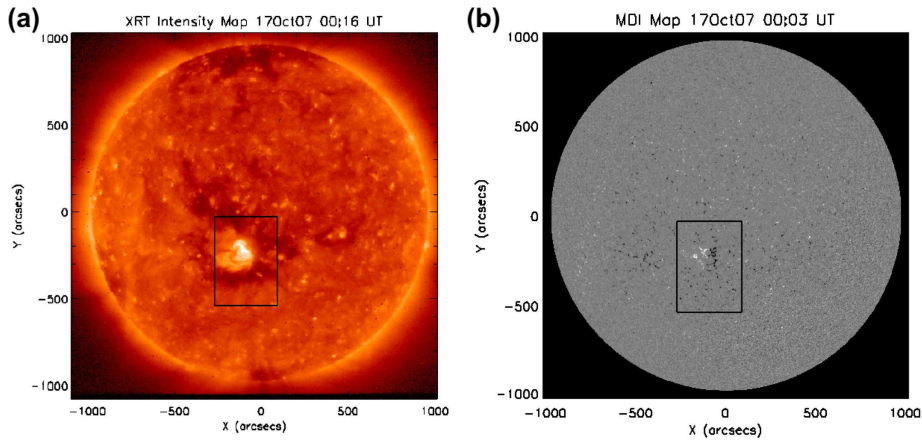}
\caption{An active region embedded in a coronal hole on 17 October 2007. (a) \emph{Hinode}/XRT (Thin\_Al\_mesh) full-disk image, with the FOV of the EIS images in Figure~\ref{Fig:Obs2} given by the black box. (b) SOHO/MDI full-disk magnetic map (white: positive polarity, black: negative polarity).}\label{Fig:Obs}
\end{figure}
A small mature AR surrounded by a CH was observed near disk centre on 17 October 2007 by the \emph{Hinode} and SOHO satellites, as shown in Figure~\ref{Fig:Obs}. Full-disk magnetograms from SOHO/Michelson Doppler Imager (MDI), corrected for line-of-sight distortion, identified the AR as orientated approximately East-West with a leading negative polarity and the CH was comprised of predominantly negative magnetic flux with an approximate strength of $-19.5\ \mathrm{gauss}\ \mathrm{(G)}$. We present data obtained by EIS showing the structure of the AR and velocity evidence of the outflows at the periphery of the AR.

The EIS study employed for this observation (study ID $205$) is composed of $24$ emission lines from ions formed over coronal temperatures of $0.3-5\ \mathrm{MK}$ and He {\sc ii}. A raster scan using the $2^{\prime\prime}$ slit and consisting of $180$ scan positions, with exposure time of $45$ seconds per position, was performed with EIS from 00:26 to 02:38 UT. The EIS field-of-view (FOV) for this study is $360^{\prime\prime}\times512^{\prime\prime}$ and covered the AR and a section of the surrounding CH. EIS data reduction was carried out using standard EIS procedures included in SolarSoft. The raw spectral data were corrected for dark current, hot and warm pixels, and cosmic rays. Doppler velocities were determined by fitting a simple Gaussian function to the calibrated spectra in order to obtain the line centre for each spectral profile. The velocities were further corrected by removing instrumental effects including slit tilt and orbital variation with standard routines.

\begin{figure}
\includegraphics[width=1.00\textwidth]{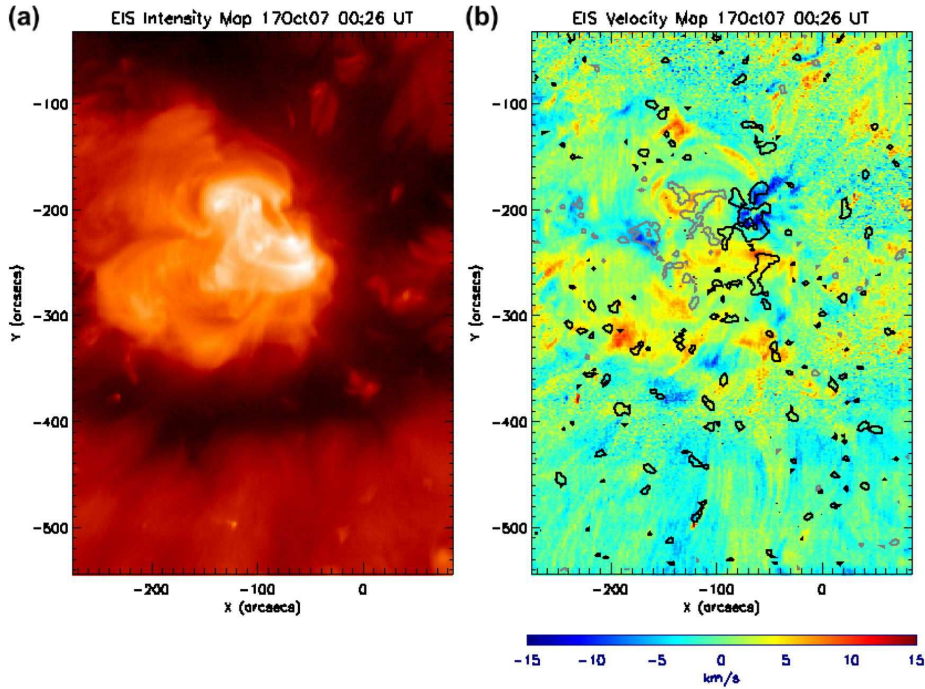}
\caption{EIS observations of the active region embedded in a coronal hole on 17 October 2007. (a) Fe {\sc xii} $195\mathrm{\AA}$ intensity map. (b) Doppler map derived from Fe {\sc xii} $195\mathrm{\AA}$, overlaid with $\pm 50\ \mathrm{G}$ contours of the line-of-sight magnetic field where gray/black is for positive/negative values.}\label{Fig:Obs2}
\end{figure}
The sigmoidal loop structure of the AR and the surrounding CH are easily identifiable in the Fe {\sc xii} $195\mathrm{\AA}$ intensity image $(1.3\ \mathrm{MK})$ shown in Figure~\ref{Fig:Obs2}~(a). Less bright loops at the solar north-eastern and south-eastern edge give the AR a sea-anemone appearance and are anticipated to have resulted from reconnection between the AR and surrounding CH field \cite{Baker09a}. The sea-anemone shape is characteristic of bipoles emerging into pre-existing magnetic environments with a dominant polarity, such as CHs \cite{Shibata94}.

Data counts within the CH were sufficient for spectrum fitting and, therefore, we were able to analyse the velocity structure of the entire FOV. A Doppler velocity map made from the Fe {\sc xii} $195\mathrm{\AA}$ line is shown in Figure~\ref{Fig:Obs2}~(b). The closed loops of the AR are clearly red-shifted, with downflows of up to $15\ \mathrm{km}\ \mathrm{s}^{-1}$, while blue-shifted outflows of magnitude $5-15\ \mathrm{km}\ \mathrm{s}^{-1}$ are located at the periphery of the AR. As described in Section~\ref{Sec:Intro} for ARs surrounded by quiet Sun, the proximity of the red and blue-shifted regions to each other generates steep gradients in the line-of-sight velocity across their interface and the outflowing plasma is located over photospheric monopolar regions of significant field strength, as illustrated in Figure~\ref{Fig:Obs2}~(b).

Several other velocity features are also identifiable in Figure~\ref{Fig:Obs2}~(b). The CH has predominantly blue-shifted velocities of magnitude $5-10\ \mathrm{km}\ \mathrm{s}^{-1}$, in line with the findings of Tu \textit{et al.} \shortcite{Tu05}, and the quiet Sun loops have velocities of $-3$ to $3\ \mathrm{km}\ \mathrm{s}^{-1}$, with red-shifted regions again corresponding to plasma draining in the loops. Along the border between the CH and quiet Sun, blue-shifted regions with velocities of magnitude $5\ \mathrm{km}\ \mathrm{s}^{-1}$ can be seen and are anticipated to occur as a result of continual interchange reconnection between the open magnetic field of the CH and the loops of the quiet Sun \cite{Crooker02,Baker07,Baker09a}.

Despite the AR being in a CH, the locations and properties of the outflows detailed above are consistent with the outflows described in Section~\ref{Sec:Intro} for ARs surrounded by quiet Sun. The outflows of the AR-CH complex are observed to persist for several days. Furthermore, comparison of the EIS intensity and velocity maps in Figure~\ref{Fig:Obs2} shows that the outflows emanate from regions of lower radiance than the sigmoidal AR while overlays of the line-of-sight photospheric magnetic field from MDI confirm the co-alignment of strong mono-polarity concentrations and the outflow locations. Thus, despite the AR being in a CH, the locations and properties of the outflows detailed above are consistent with the outflows described in Section~\ref{Sec:Intro} for ARs surrounded by quiet Sun and, hence, we deduce that outflows of a similar nature can be instigated at the periphery of an AR regardless of whether it is encompassed by quiet Sun or a CH.

The simplicity of this AR-CH complex provides an opportunity to probe the driving mechanism of the outflows using three-dimensional simulations. The simulation setup and results will be presented in Sections~\ref{Sec:Sim} and \ref{Sec:Res}, respectively.

\section{Simulation Setup}\label{Sec:Sim}
In order to investigate the origin of the outflows observed at the edges of the AR in a CH and their acceleration mechanism, we have performed a numerical simulation that emulates the development and evolution of such an AR. Several simplifications to the observed scenario have been made. Firstly, we model only the AR and the CH and neglect the quiet Sun regions and their associated flows. Secondly, we neglect to include the ubiquitous outflows of the CH, choosing to focus on the generation of the outflows at the periphery of the AR.

\subsection{\sc{Model}}\label{Sec:model}
\begin{figure}\centering
\includegraphics[width=1.0\textwidth]{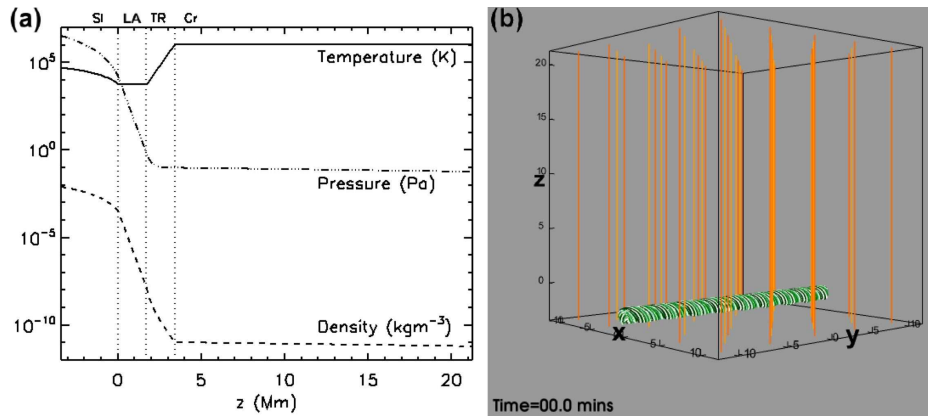}
\caption{Initial conditions of the simulation domain. (a) Profiles of temperature (solid), gas pressure (triple dot-dashed) and density (dashed) in height, with the solar interior (SI), low atmosphere (LA), transition region (TR) and corona (Cr) indicated by vertical dotted lines. (b) Selection of magnetic fieldlines of the CH (orange and yellow) and flux tube (white and shades of green).}\label{Fig:Sim_atmo}
\end{figure}

The domain comprises four horizontal layers representing the top of the solar interior ($-3.4 \leq z \leq 0.0\ \mathrm{Mm}$), the low atmosphere ($0.0 \leq z \leq 1.7\ \mathrm{Mm}$), the transition region ($1.7 \leq z \leq 3.4\ \mathrm{Mm}$) and the low corona ($3.4 \leq z \leq 21.25\ \mathrm{Mm}$). Each layer is defined by a distinct temperature characteristic, illustrated in Figure~\ref{Fig:Sim_atmo}~(a), from which the gas pressure and density are obtained under the assumption of hydrostatic equilibrium. The solar interior is chosen to have a linear temperature profile, with a gradient such that the region is marginally stable to the onset of convection. The low atmosphere, representing the photosphere and chromosphere, and corona are isothermal regions of $5600\ \mathrm{K}$ and $1.1\ \mathrm{MK}$, respectively. The transition region is sandwiched by the two isothermal regions and is characterised by a rapid increase in temperature. The whole of the domain is endowed with a vertical magnetic field of $-19.5\ \mathrm{G}$, shown in Figure~\ref{Fig:Sim_atmo}~(b). This domain represents the simplified conditions of a CH, in which all field is taken to be uniform and open to interplanetary space and the ubiquitous $10-20\ \mathrm{km}\ \mathrm{s}^{-1}$ outflow of the CH is neglected.

The AR in the CH will develop from the emergence of a flux tube through the solar surface ($z=0.0\ \mathrm{Mm}$). The magnetic field of the flux tube, also shown in Figure~\ref{Fig:Sim_atmo}~(b), is prescribed by $\textbf{B}=\left(B_r,B_\theta,B_y\right)$, where
\begin{eqnarray}
B_r&=&0,\\
B_\theta&=&\alpha r B_y,\\
B_y&=&B_0 e^{-r^2/R^2}.
\end{eqnarray}The maximum field strength of the tube is given by $B_0=5.2\times 10^3\ \mathrm{G}$, the radius at which the axial field strength falls by a factor of 1/e is set at $R=0.425\ \mathrm{Mm}$ and the degree of rotation of the fieldlines about the axis is taken as $\alpha=0.37\times 2\pi$ for each $1\ \mathrm{Mm}$ length in the axial direction. The axis of the flux tube is set in the solar interior at $-1.7\ \mathrm{Mm}$. At $t=0$, the tube is chosen to be in radial force balance with the surrounding medium in order to prevent a sudden radial expansion or compression of the tube once the simulation begins. Additionally, the central portion of the tube, at $y=0$, is chosen to be in thermal equilibrium with the environment but the temperature in the tube decreases away from this region. The variation in temperature results in the central portion of the tube being the most buoyant, with decreasing buoyancy towards the ends. The total buoyant length of the tube in the axial direction is $\approx 17\ \mathrm{Mm}$. The differential buoyancy aids the development of an $\Omega$-shaped loop along the tube's axial length that will rise towards the solar surface when the simulation commences.

The domain models the region $-11.9\ \mathrm{Mm}\leq x \leq 11.9\ \mathrm{Mm}$, $-11.9\ \mathrm{Mm}\leq y \leq 11.9\ \mathrm{Mm}$ and $-3.4\ \mathrm{Mm}\leq z \leq 21.25\ \mathrm{Mm}$ using $280\times280\times240$ gridpoints of uniform spacing. The boundaries of the domain are periodic in $x$ and $y$ and closed in $z$ and no damping zones are imposed.

\subsection{\sc{Numerical method}}
The model domain described in Section~\ref{Sec:model} is advanced in time using the Lagrangian remap scheme Lare3d \cite{Arber01}, which numerically solves the following time-dependent, resistive, dimensionless MHD equations:
\begin{eqnarray}
\frac{\partial \rho}{\partial t} &=& -\nabla\cdot\left(\rho\textbf{v}\right),\label{Equ:MHD1}\\
\frac{\partial \rho \textbf{v}}{\partial t} &=& -\nabla\cdot\left(\rho\textbf{v}\textbf{v}\right)+\left(\nabla\times\textbf{B}\right)\times\textbf{B}-\nabla p+\rho\textbf{g},\label{Equ:MHD2}\\
\frac{\partial \textbf{B}}{\partial t} &=& \nabla\times\left(\textbf{v}\times\textbf{B}\right)+\eta\nabla^2\textbf{B},\label{Equ:MHD3}\\
\frac{\partial \rho \epsilon}{\partial t} &=& -\nabla\cdot\left(\rho\epsilon\textbf{v}\right) -p\left(\nabla\cdot\textbf{v}\right) +Q_\mathrm{joule}+Q_\mathrm{shock},\label{Equ:MHD4}
\end{eqnarray}with density $\rho$, time $t$, velocity $\textbf{v}$, magnetic field $\textbf{B}$, gas pressure $p$, acceleration due to gravity $\textbf{g}=-g\ \textbf{z}$, resistivity $\eta$, specific energy density $\epsilon = p/(\gamma-1)\rho$, joule heating $Q_\mathrm{joule}$ and viscous heating at shocks $Q_\mathrm{shock}$. The ratio of specific heats, gravitational acceleration and resistivity are taken to be constants with dimensionless values $\gamma=5/3$, ${g=1.0}$ and $\eta=0.001$, respectively. Heat conduction and radiative effects are neglected.

The values returned from Equations~$\left(\ref{Equ:MHD1}\right)$--$\left(\ref{Equ:MHD4}\right)$ are made dimensional using the following solar photospheric values: pressure scale height $H_\mathrm{ph}=1.7\times10^5\ \mathrm{m}$, time $t_\mathrm{ph}=25.0\ \mathrm{s}$, velocity $v_\mathrm{ph}=6.8\times10^3\ \mathrm{m}\ \mathrm{s}^{-1}$, density $\rho_\mathrm{ph}=3.0\times10^{-4}\ \mathrm{kg}\ \mathrm{m}^{-3}$, pressure $p_\mathrm{ph}=1.2\times10^4\ \mathrm{Pa}$, temperature $T_\mathrm{ph}=5.6\times10^3\ \mathrm{K}$, magnetic field $B_\mathrm{ph}=1.3\times10^3\ \mathrm{G}$, surface gravity $g_\mathrm{ph}=2.7\times10^2\ \mathrm{m}\ \mathrm{s}^{-1}$ and mean atomic weight $\tilde{\mu}=1.0$. All values discussed within this paper will be dimensional unless otherwise stated.

\section{Results}\label{Sec:Res}
The initial stages of the experiment are thoroughly documented in previous literature and are given here in brief for completeness only. The reader is directed to Murray \textit{et al.} \shortcite{Murray06} and references therein for detailed information regarding the emergence process.

When the simulation commences, the central portion of the flux tube begins rising towards the solar surface and the tube adopts an $\Omega$-shape. The fall in density of the background plasma in the solar interior with height means that, whilst still rising in this medium, the tube eventually looses all buoyancy. Nevertheless, it continues to rise due to the marginally stable nature of the solar interior to the onset of convection. Upon reaching the solar surface the tube can no longer continue to proceed under the same mechanism since the atmosphere is strongly subadiabatic. In fact, the flux tube must succumb to a magnetic buoyancy instability before it can proceed further. In this simulation, the flux tube is of sufficient magnetic strength and twist that the condition for the occurrence of this instability is immediately satisfied upon reaching the base of the atmosphere. The magnetic field of the tube then expands rapidly into the atmosphere, both vertically and horizontally, driven by a magnetic pressure gradient to form an AR.

While plasma is draining in the loops of the AR, we find that outflows of plasma occur at the AR's periphery, with maximum vertical velocities of $45\ \mathrm{km}\ \mathrm{s}^{-1}$ and lasting for only a few minutes. In Section~\ref{Sec:Prop} we detail the origin and properties of the outflowing plasma and in Section~\ref{Sec:Drive} we discuss the mechanism driving the outflows.

\begin{figure}\centering
\includegraphics[width=1.0\textwidth]{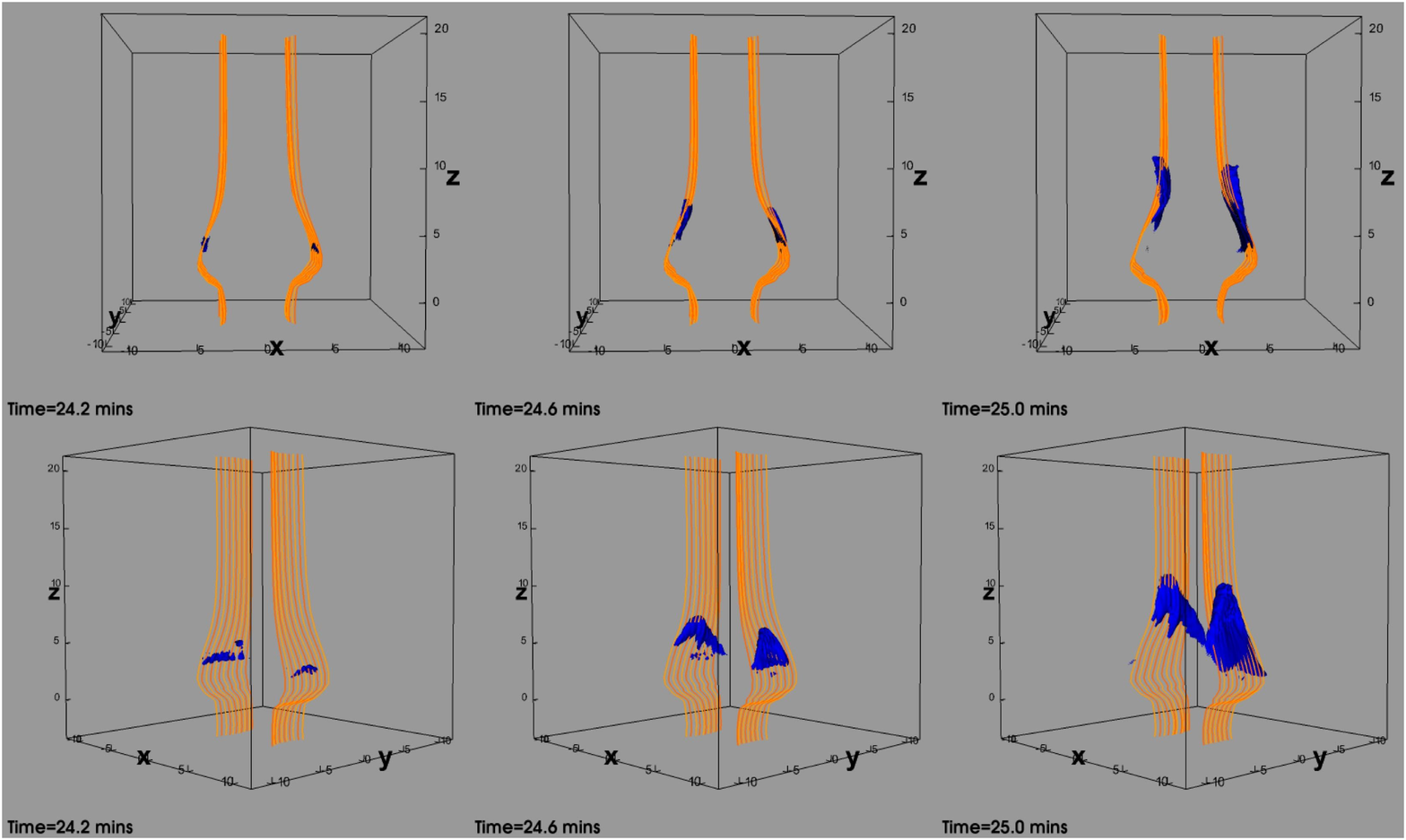}
\caption{Blue vertical velocity isosurfaces of $15\ \mathrm{km}\ \mathrm{s}^{-1}$ showing the evolution of the outflowing plasma at $24.2$, $24.6$ and $25.0\ \mathrm{mins}$ into the simulation from two different perspectives. A selection of coronal hole fieldlines threading through the outflow are shown in yellow and orange.}\label{Fig:Sim_Bevo}
\end{figure}
\subsection{\sc{Origin and Properties of the Outflows}}\label{Sec:Prop}
The outflowing plasma originates from the CH region immediately neighbouring the AR. It rises from the base of the transition region with a mostly vertical trajectory, through the corona to the upper boundary of the simulation domain, as illustrated in Figure~\ref{Fig:Sim_Bevo}. The temperature and density of the outflowing material are $\approx 1.4\ \mathrm{MK}$ and of order $10^{-11}\ \mathrm{kg}\ \mathrm{m}^{-3}$, respectively. Given the source of the outflowing plasma, it is unsurprising that the temperature and density of the material are more representative of the initial conditions of the CH than the developing AR. As detailed in Section~\ref{Sec:Intro}, this is a well documented characteristic of outflows observed at the periphery of ARs. Furthermore, the lose of mass from the neighbouring CH field will have little impact on the AR itself, another reported fact from the observations.

\begin{figure}\centering
\includegraphics[width=1.0\textwidth]{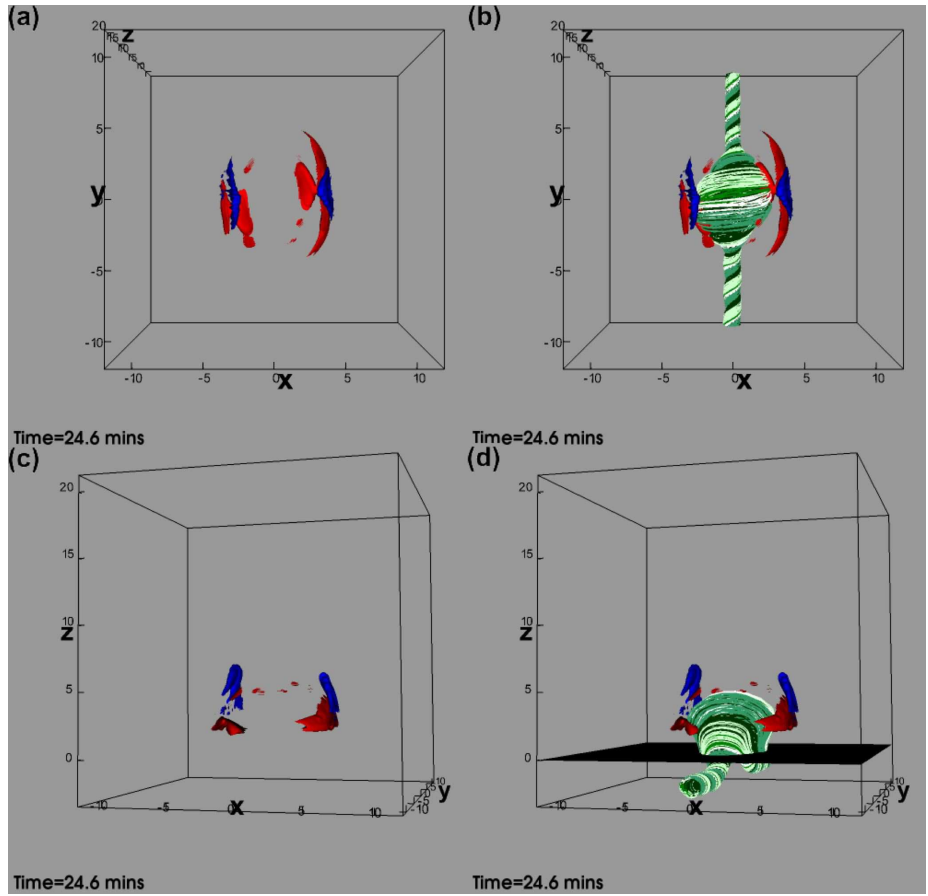}
\caption{Isosurfaces of vertical velocity for the outflowing plasma at the periphery of the active region (blue) and draining plasma in the active region's loops (red) from two different perspectives, with and without a selection of the active region's fieldlines (white and shades of green) at $24.6\ \mathrm{mins}$ into the simulation. The isosurfaces are set at levels of $15$ and $-15\ \mathrm{km}\ \mathrm{s}^{-1}$ for the outflowing and draining plasmas, respectively. The horizontal plane in (d) gives the location of the solar surface.}\label{Fig:Sim_RBshift}
\end{figure}

Figure~\ref{Fig:Sim_RBshift} shows the location of the outflowing plasma of the CH in context with the draining plasma in the AR loops. The outflows sit immediately next to the locations of draining, which is a distinctive feature in the EIS velocity map of the AR-CH complex (Figure~\ref{Fig:Obs2}~(b)). The two different flow regimes are separated by a separatrix layer, an invisible boundary across which the connectivity of the magnetic field changes. The proximity of these oppositely directed flows to each other results in steep gradients being generated across the separatrix layer.

\begin{figure}\centering
\includegraphics[width=0.75\textwidth]{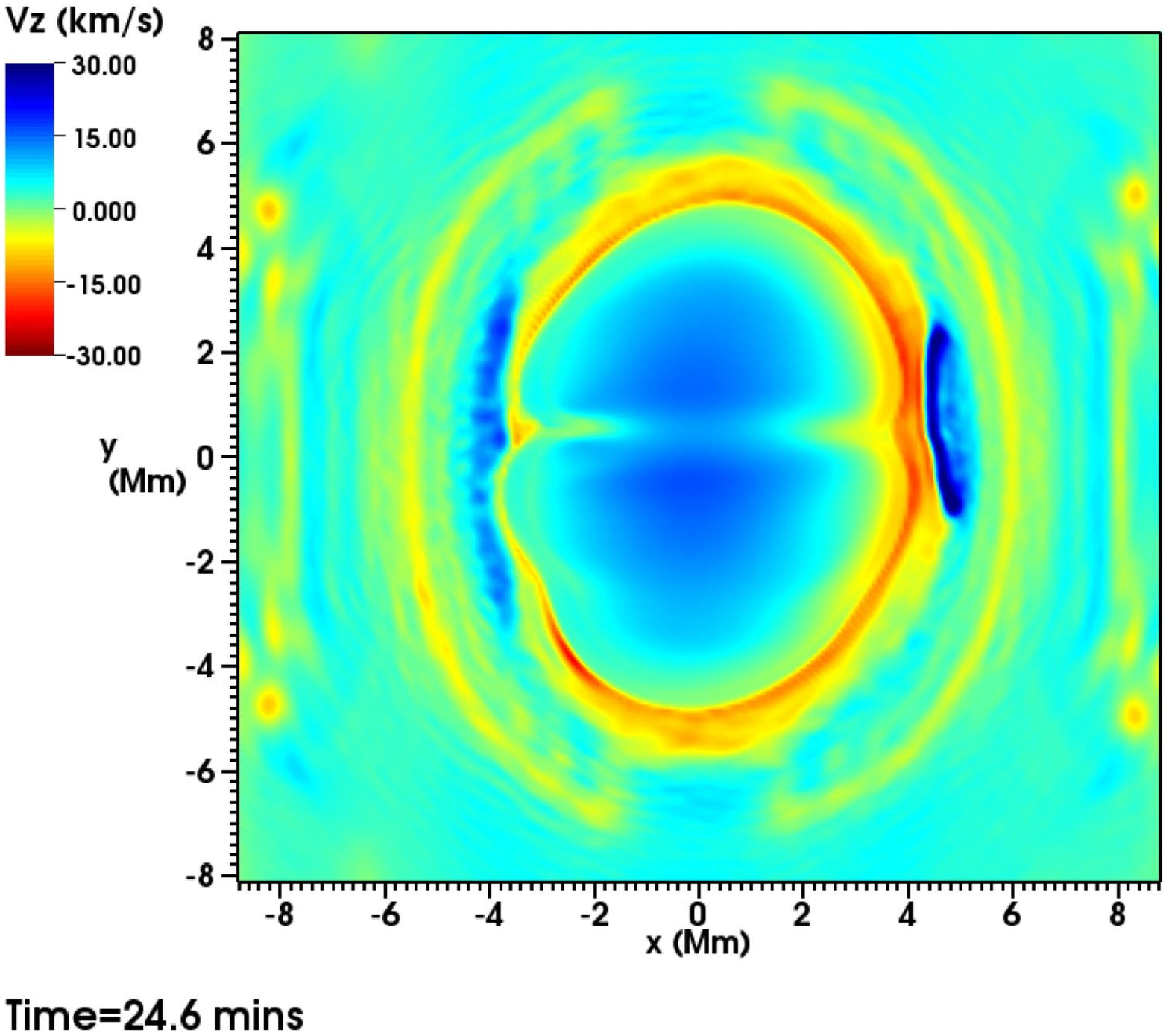}
\caption{Horizontal slice through the domain at $z=4.5 \ \mathrm{Mm}$ of vertical velocity showing the outflowing plasma (blue) of the coronal hole located next to the draining plasma (red) of the active region's loops at $24.6\ \mathrm{mins}$ into the simulation.}\label{Fig:Sim_RBshift_map}
\end{figure}

A map of vertical velocity at a coronal height of $z=4.5\ \mathrm{Mm}$, shown in Figure~\ref{Fig:Sim_RBshift_map}, again emphasises the proximity of the draining plasma at the edge of the AR and the rising plasma in the neighbouring CH. The centre of the AR is also dominated by an upflow. This upflow is perpendicular to the direction of the AR's loops and occurs while the magnetic field that forms the AR is transported into the atmosphere during the development stages. No similar central upflow is visible in the EIS map presented in Section~\ref{Sec:Obs}. The absense may be due to the fact that the AR is at a mature stage of its lifecycle and will have long since finished any dramatic emerging activity and accompanying vertical expansion or the upflow may simply be masked by faster downflows along the line-of-sight.
\begin{figure}\centering
\includegraphics[width=0.75\textwidth]{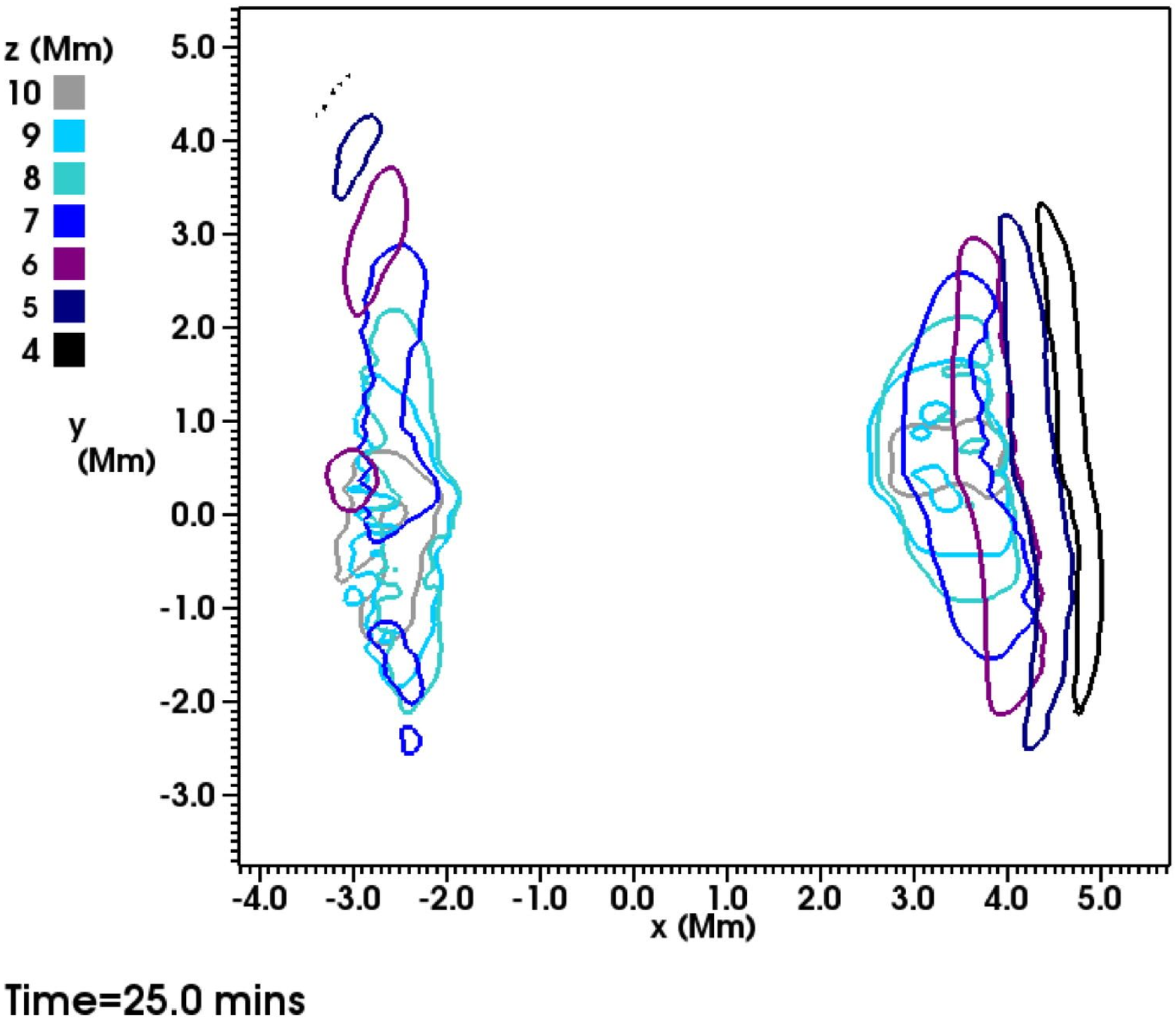}
\caption{$15\ \mathrm{km}\ \mathrm{s}^{-1}$ vertical velocity contours of the outflows in the $x-y$ plane from $4\ \mathrm{Mm} \le z \le 10\ \mathrm{Mm}$ at $1\ \mathrm{Mm}$ spaced intervals in height projected onto a single $x-y$ plane at $25.0\ \mathrm{mins}$ into the simulation.}\label{Fig:Sim_Bshift_str}
\end{figure}

As explained above, the outflowing plasma is accelerated from its source site and channelled along the CH magnetic field it is essentially ``frozen'' to. Hence, the shape of the CH field is exceedingly important in determining the path and structure of the outflows. As the AR expands during its development stages, it pushes the CH field out of its way. In the lower atmosphere the footpoints of the CH field neighbouring the AR become compacted but in the upper atmosphere where the AR does not encroach the CH field remains in its initial decompressed state. Hence, the CH field takes on a new shape, whereby it curves around the AR's loops, as visible in Figure~\ref{Fig:Sim_Bevo}, and expands slightly with height. Thus, as the outflows rise they follow the curve and expansion of the CH field. This is depicted in Figure~\ref{Fig:Sim_Bshift_str} where horizontal slices have been taken through the atmosphere at $1\ \mathrm{Mm}$ intervals and the $15\ \mathrm{km}\ \mathrm{s}^{-1}$ vertical velocity contour of each slice has been plotted. The origin of the outflow is a thin elongated region but, at greater heights in the atmosphere, the outflows have migrated horizontally to locations over the AR as they follow the curve of the CH field during their rise. The outflows also become less elongated and more elliptical in shape as they travel along the CH magnetic field which diverges slightly with height. The structure of the outflows with increasing height in the simulation is similar to that described by Baker \textit{at al.} \shortcite{Baker09} for increasing temperature. Under the assumption of increasing temperature with height, the two results appear to be consistent with each other. However, the migration aspect does not appear to be an obvious feature in their example and could be due to the neighbouring field being more divergent with height in their case.

\begin{figure}\centering
\includegraphics[width=0.75\textwidth]{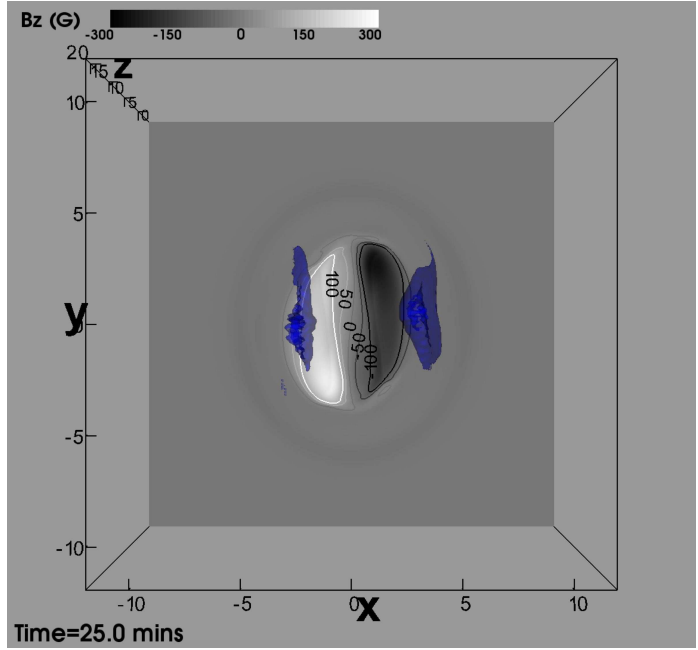}
\caption{$15\ \mathrm{km}\ \mathrm{s}^{-1}$ transparent blue isosurfaces of vertical velocity for the outflows and a horizontal photospheric plane $(z=0.8\ \mathrm{Mm})$ of vertical magnetic field with contours ($0$, $\pm50$, $\pm100\ \mathrm{G}$) shown at $25.0\ \mathrm{mins}$ into the simulation. The outflows on both sides migrate towards the active region's centre as they rise up through the atmosphere.}\label{Fig:Sim_Bshift_magf}
\end{figure}

The migration of the outflows over the AR as they rise has an important consequence when making comparisons between their location and photospheric quantities. For example, although the outflows are accelerated from a weak magnetic field region in the CH, as they rise the curvature of the CH field brings them directly over the magnetically stronger monopolar footpoints of the AR as shown in Figure~\ref{Fig:Sim_Bshift_magf}. The alignment of projected outflows with strong photospheric monopolar concentrations is a prominent characteristic of the outflows observed in the AR-CH complex, shown in Figure~\ref{Fig:Obs}~(b). Sections of the line formed by the intersection of the separatrix surface between the AR and the CH with the solar surface will also fall under the outflows as a result of the latter's migration.

In summary, the simulated development of an AR in a CH has resulted in the generation of outflows at the periphery of the AR with a maximum vertical velocity of $45\ \mathrm{km}\ \mathrm{s}^{-1}$. The outflows are accelerated in and along the CH magnetic field immediately neighbouring the AR. The characteristics of the outflows (temperature, density and structure) and their location with respect to the draining plasma of the AR's loops and photospheric monopolar magnetic concentrations are compatible with the descriptions of the outflows given in Section~\ref{Sec:Obs} for the AR-CH complex and in Section~\ref{Sec:Intro} for ARs surrounded by quiet Sun. The maximum velocity of the outflows generated in the simulation is faster than that observed for the AR in the CH but the photospheric magnetic field is also stronger in the former and, therefore, appears to fit with the finding of Doschek \textit{et al.} \shortcite{Doschek08} that magnetically stronger ARs are associated with faster outflows. The limitations of the simulation in generating outflows consistent with observations will be considered in Section~\ref{Sec:Dis}, once the driving mechanism has been fully explained in Section~\ref{Sec:Drive}.

\subsection{\sc{Driving Mechanism of the Outflows}}\label{Sec:Drive}
Examination of the forces at work in the system reveals that the outflows occur simply as a consequence of the AR expanding horizontally.

\begin{figure}\centering
\includegraphics[width=1.0\textwidth]{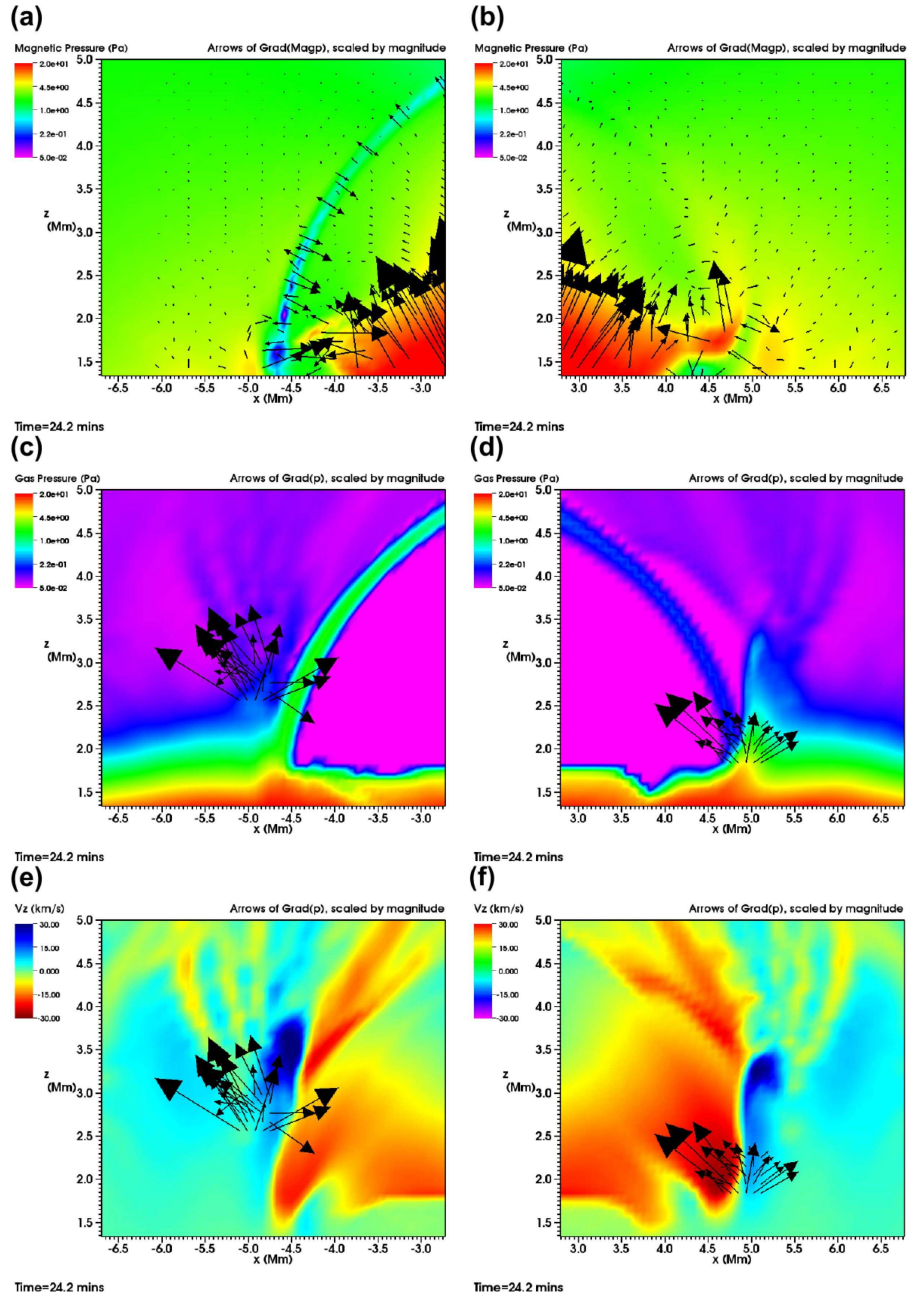}
\caption{Vertical slices through a subspace of the domain at the left-handside (a, c, e) and right-handside (b, d, f) interfaces between the active region and coronal hole at $24.2\ \mathrm{mins}$ into the simulation. Top: contours of magnetic pressure overlaid with arrows indicating the magnitude and direction of the magnetic pressure gradient. Middle: contours of gas pressure overlaid with a selection of arrows indicating the magnitude and direction of the gas pressure gradient at the acceleration point, Bottom: contours of vertical velocity overlaid with a selection of arrows indicating the magnitude and direction of the gas pressure gradient in the vicinity of the outflows. The slices are taken at $y=-0.4\ \mathrm{Mm}$ and $y=0.3\ \mathrm{Mm}$ for the left-handside and right-handside, respectively.}\label{Fig:Sim_Force}
\end{figure}

As explained at the beginning of Section~\ref{Sec:Res}, the vertical and horizontal expansion of the AR is driven by a magnetic pressure gradient. The flux tube from which the AR develops is the strongest magnetic entity in the domain, as shown in Figure~\ref{Fig:Sim_Force}(a) and (b), and the associated magnetic pressure gradient dominates over the combined gradients in the gas pressure and magnetic pressure of the CH environment, thus driving the AR's expansion.

It is the horizontal expansion of the AR that has the greatest impact on the surrounding CH environment. The horizontal expansion of the AR compresses and deforms the vertical CH magnetic field, as discussed in Section~\ref{Sec:Prop}, generating Lorentz forces in the latter. The horizontal expansion of the AR also compresses the plasma amassed in the CH field, which results in enhanced gas pressure gradients and gravitational forces there. The compressive effects are evident in Figure~\ref{Fig:Sim_Force}(a)-(d), where enhanced magnetic pressure and gas pressure can be seen in the CH region at approximately $x=\pm5.0\ \mathrm{Mm}$. Of the three types of forces generated in the CH field, only the gas pressure gradients act in a direction parallel to the magnetic field and have components acting in the positive vertical direction. Hence, only the gas pressure gradients can drive the outflows up and along the CH field. Figure~\ref{Fig:Sim_Force}(e) and (f) show the co-spatial agreement of the gas pressure gradients and the origins of the accelerated plasma. A gas pressure gradient acting parallel to the CH field is always present in the domain but at $t=0$ it is exactly balanced by a gravitational force. However, after the development of the AR, the enhanced element of the parallel gas pressure gradient is not fully counteracted and acts to accelerate the plasma in the CH field upwards, thus producing outflows.

Under this driving mechanism, the velocity of the outflowing plasma is dependent upon the size of the imbalance in the components of the enhanced gas pressure gradient and gravitational force that are parallel to the CH field, with larger imbalances giving faster velocities. The size of the enhanced gas pressure gradient will in turn depend upon the size of the force driving the AR expansion, with larger expansion forces resulting in greater compression and larger gas pressure gradients. Larger expansion forces will be caused by a weaker CH field or/and a stronger AR. The speed of the horizontal expansion of the AR should be an indicator of the size of the expansion driving force and, hence, should be positively linked to the speed of the outflows. This hypothesis is left for testing in a future study.


Along the left-handside interface between the AR and CH, a region of high gas pressure and low magnetic pressure is generated, as visible in Figure~\ref{Fig:Sim_Force}(a) and (c). These are classic characteristics of a current sheet, which develops due to the practically oppositely orientated magnetic fields of the AR and CH at the left-handside interface. Conversely, the magnetic fields of the AR and CH at the right-handside interface are directed in approximately the same direction and, hence, no current sheet develops there. The existence of the current sheet is important for the long-term evolution of the system since reconnection ultimately set in but it has only minor effects on the outflow at the left-handside of the AR. For example, the left-handside outflow is accelerated from a point that is $0.5\ \mathrm{Mm}$ higher in the atmosphere than the right-handside outflow, as shown in Figure~\ref{Fig:Sim_Force}(e) and (f).

As stated in Section~\ref{Sec:Res}, the outflows only persist for a couple of minutes. Reconnection sets in between the AR and CH fields and irreparably alters the forces in the CH field that contribute to the driving mechanism at both the reconnecting and non-reconnecting interfaces. This brings a halt to the acceleration of further plasma. For technical reasons, the three-dimensional simulation is curtailed at this point but 2.5-dimensional simulations, using the Lare2.5D code with varying flux tube and CH field strengths, reveal that oscillatory reconnection sets in after the outflows have occurred. As a result of this reconnection, the AR develops a two-dimensional sea-anemone structure and the flux systems eventually settle into an equilibrium state \cite{Murray09}. There are no imposed drivers in the system to perturb the domain from this equilibrium and, hence, continuing these 2.5-dimensional simulations reveals no further dynamics. 

\section{Discussion}\label{Sec:Dis}
The simulation we have performed clearly shows the existence of outflows that originate in the magnetic field of the CH immediately neighbouring the AR. These outflows are accelerated simply by the horizontal expansion of the AR. The characteristics of the outflows are in line with those reported for outflows observed at the edges of the AR in the CH, presented in Section~\ref{Sec:Obs}, and ARs surrounded by quiet Sun, detailed in Section~\ref{Sec:Intro}. It is, therefore, possible that the acceleration mechanism present in the simulation could be, at least partially, responsible for generating the observed outflows at the periphery of ARs surrounded by both CHs and quiet Sun. In this section, we consider the plausibility of this driving mechanism for generating the outflows observed by EIS at the edges of ARs and the limitations of the simulation.

In the simulation, the expansion of the AR's loops is an integral part of AR development and a natural consequence of the emergence of the strong magnetic field from the solar interior. Yet the AR considered in Section~\ref{Sec:Obs} is a mature AR, as are the ARs studied by EIS to date that have outflows at their periphery. These mature ARs will have long since undergone such dramatic emergence and subsequent expansion as that in the simulation. However, it has been observed that expansion is a general feature of established ARs \cite{Uchida92}, with velocities of a few to a few tens of $\mathrm{km}\ \mathrm{s}^{-1}$. Expansion may occur as a result of tether-cutting reconnection along the polarity inversion line of the AR or continuous dispersion of the AR's footpoints. Hence, it is entirely plausible that the acceleration mechanism described in this paper is occurring at the edges of ARs as they expand horizontally.

As explained in Section~\ref{Sec:Drive}, it is anticipated that the expansion speed of the AR will be positively related to the velocity of the outflows. The speed of horizontal expansion in the simulation lies comfortably within the range given by Uchida \textit{et al.} \shortcite{Uchida92}. Thus, if the expansion acceleration mechanism were at work in ARs then we would expect the outflow velocity of the simulation and observations to be consistent with each other. This is indeed the case and is another indication that the expansion acceleration mechanism could plausibly be producing the outflows observed at the edges of ARs.

The anticipated positive relationship between AR expansion speed and outflow velocity would also explain why ARs with stronger magnetic footpoints have been observed to have faster outflows at their periphery. In emergence simulations it has been found that the magnetic strength of AR footpoints is positively correlated to the strength of the magnetic field that emerged from the solar interior to form the AR. Stronger magnetic fields expand faster into the atmosphere due to their associated higher magnetic pressure \cite{Murray06} and, hence, should drive faster outflows if the rationale linking expansion and outflow speeds given in Section~\ref{Sec:Drive} is correct.

The magnetic field neighbouring the AR in the simulation is loosely based on that of a CH, namely vertical and of uniform strength. However, in reality, the magnetic field of the quiet Sun and CHs consists of multi-scale loops and open field, covering a spectrum of orientations at points of contact with the loops of ARs. But can we expect the expansion acceleration mechanism to operate in the neighbouring magnetic field if it is comprised of loops? The only necessary elements for the expansion mechanism to operate are a horizontally expanding AR and a compressible neighbouring magnetic field with a vertical component $(B_z\ne0)$ at the potential acceleration site. Providing these elements are present, we believe that flows should be generated via horizontal AR expansion regardless of the closed or open state of the neighbouring field. In the case of neighbouring loops, the flows will no longer be outflows since they will be confined by the magnetic field and would be siphon flows. The results of the simulation show that outflows can be produced irrespective of whether the fields are aligned for reconnection or not and, thus, we also believe that the orientation of the neighbouring magnetic field to the AR's loops will be of little hindrance to the generation of flows. Further simulations will be required to fully verify these assertions.

The generation of the outflows for only a short period of time is a significant limitation of the simulation. As explained in Section~\ref{Sec:Drive}, the reconnection between the AR and CH fields irreparably alters the forces in the CH field that contribute to the driving mechanism, bringing a halt to the acceleration of further plasma. The results from 2.5-dimensional simulations reveal that oscillatory reconnection takes the system to an equilibrium state and, therefore, no further outflows are generated once the reconnection ceases \cite{Murray09}. If the anemone-shaped AR could be perturbed from this equilibrium it may well be possible to drive further outflows. The disturbance would need to lead to the AR undergoing further expansion that compresses the neighbouring magnetic field. Additionally, it is anticipated that continual expansion of the AR in the absence of reconnection would result in outflows lasting for longer periods of time, a necessary requirement in order to be in line with the persistency shown in observations.

\section{Conclusions}\label{Sec:Con}
We have performed a three-dimensional simulation of an AR developing in a CH. The simulation is motivated by observations of an AR embedded in a CH on 17 October 2007 by the \emph{Hinode} satellite. Velocity maps constructed from data taken by the EIS instrument onboard \emph{Hinode} reveal distinctive outflows of plasma situated next to the draining loops of the AR. These outflows are a common feature of velocity maps with ARs surrounded by quiet Sun and several theories have been proposed to explain them. We have used the simulation to investigate the occurrence of these outflows in more detail.

We find that outflows are indeed generated next to the draining plasma in the AR's loops, with the two flow regimes separated by a separatrix surface. The outflows originate in the CH field immediately neighbouring the AR, accelerated simply by the horizontal expansion of the AR. The outflows have vertical velocities of up to $45\ \mathrm{km}\ \mathrm{s}^{-1}$ and temperatures and densities in the region of the initial conditions of the CH. The outflows travel along the CH magnetic field and, hence, the shape of the CH field is important for determining the structure of the outflows. CH field that curves around the AR's loops will transport the outflowing material over the AR and, therefore, projection of the outflows over photospheric monopolar magnetic concentrations and separatrix layers occurs. Though the acceleration site is an elongated and narrow region, the expansion of the CH field results in the column of outflowing plasma becoming more circular and extended with height.

These characteristics are in line with those of the outflows in the AR-CH complex described in Section~\ref{Sec:Obs} and outflows observed at the periphery of ARs surrounded by quiet Sun summarised in Section~\ref{Sec:Intro}. Thus, the acceleration of plasma due to the horizontal expansion of the AR is a viable mechanism for generating the correct type of outflows at the edges of ARs embedded in open field such as is found in CHs and, as discussed in Section~\ref{Sec:Dis}, we believe it is equally applicable to ARs surrounded by loops such as in the quiet Sun. Outflows generated by horizontal expansion of an AR is also a plausible mechanism since ARs are reported to continually expand even once they reach a mature stage.

Although the expansion acceleration mechanism appears to be able to adequately explain the observed outflows, it is unlikely to be working in isolation of other acceleration mechanisms such as reconnection and its by-products. As reported by Baker \textit{et al.} \shortcite{Baker09}, AR loops will be separated from any surrounding field by (quasi) separatrix layers, which are natural sites for reconnection. Expansion of the AR is one way in which the current concentrations in these layers can be increased such that they are high enough for reconnection to commence and, hence, it may be possible to simultaneously produce outflows with the mechanism described in this paper and outflows as a result of reconnection.

Finally, as stated in Section~\ref{Sec:Intro}, if the field along which the plasma is being accelerated is open into interplanetary space then the expansion acceleration mechanism could be responsible for accelerating some of the plasma that forms the slow solar wind.

\section*{Acknowledgements}
MJM and DB acknowledge financial assistance from the Science \& Technology Facilities Council (STFC) of the UK. The work of LvDG was partially supported by the European Commission through the SOTERIA Network (EU FP7 Space Science Project No. 218816). \emph{Hinode} is a Japanese mission developed and launched by ISAS/JAXA, with NAOJ as domestic partner and NASA and STFC (UK) as international partners. It is operated by these agencies in co-operation with ESA and the NSC (Norway). The computational work for this paper was carried out on the Legion linux cluster at University College London, UK.

\bibliographystyle{spr-mp-sola-cnd}
\bibliography{bibtex_WORK}

\end{article} 
\end{document}